\documentclass[journal]{IEEEtran}

\usepackage{cite}
\usepackage{amsmath,amssymb,amsfonts}
\usepackage{graphicx}
\usepackage{subfigure}
\usepackage{textcomp}
\usepackage{xcolor}

\usepackage{bm}

\usepackage[noend]{algpseudocode}
\usepackage{algorithmicx,algorithm}

\usepackage{amsmath}

\usepackage{amsmath}

\usepackage{amsthm}

\newtheorem{proposition}{Proposition}

\newtheorem{remark}{Remark}

\def\BibTeX{{\rm B\kern-.05em{\sc i\kern-.025em b}\kern-.08em
    T\kern-.1667em\lower.7ex\hbox{E}\kern-.125emX}}
    
\ifCLASSINFOpdf
\else
\fi

\allowdisplaybreaks[2]

\begin{document}

\linespread {0.99}  

\addtolength{\parskip}{0.99pt} 

% 公式间距
\makeatletter
\renewcommand\normalsize{%
\@setfontsize\normalsize\@xpt\@xiipt
\abovedisplayskip 4\p@ \@plus2\p@ \@minus5\p@
\abovedisplayshortskip \z@ \@plus3\p@
\belowdisplayshortskip 6\p@ \@plus3\p@ \@minus3\p@
\belowdisplayskip \abovedisplayskip
\let\@listi\@listI}
\makeatother

\title{Superposed IM-OFDM (S-IM-OFDM): An Enhanced OFDM for
Integrated Sensing and Communications}

\author{Zonghui~Yang,~\IEEEmembership{Graduate Student~Member,~IEEE,} Shijian~Gao,~\IEEEmembership{Member,~IEEE,}\\Xiang~Cheng,~\IEEEmembership{Fellow,~IEEE}~and Liuqing~Yang,~\IEEEmembership{Fellow,~IEEE}.
%\thanks{Z.~Yang and X.~Cheng are with the State Key Laboratory of Advanced Optical Communication Systems and Networks, School of Electronics, Peking University, Beijing 100080, China (email: yzh22@stu.pku.edu.cn; xiangcheng@pku.edu.cn).}% <-this % stops a space
%\thanks{S.~Gao was formerly with the Department of Electrical & Computer Engineering, University of Minnesota, Minneapolis, MN 55455, USA, and is now with Samsung SoC R&D Lab, San Diego, CA, 92121 (e-mail: gao00379@umn.edu).}% <-this % stops a space
%\thanks{L.~Yang is with the Internet of Things Thrust & Intelligent Transportation Thrust The Hong Kong University of Science and Technology (Guangzhou) Guangzhou, China, and also with the Department of Electronic and Computer Engineering, Hong Kong University of Science and Technology Hong Kong SAR, China (Email: lqyang@ust.hk).}% <-this % stops a space
%\thanks{Manuscript received ; revised .}
}% <-this % stops a space
%\thanks{J. Doe and J. Doe are with Anonymous University.}% <-this % stops a space
%\thanks{Manuscript received April 19, 2005; revised August 26, 2015.}}

%\markboth{IEEE TRANSACTIONS ON VEHICULAR TECHNOLOGY, 2024}%
%{Shell \MakeLowercase{\textit{et al.}}: Bare Demo of IEEEtran.cls for IEEE Journals}
%\vspace{-0.9cm}
\maketitle

\begin{abstract}
Integrated sensing and communications (ISAC) is a critical enabler for emerging 6G applications, and at its core lies in the dual-functional waveform design. While orthogonal frequency division multiplexing (OFDM) has been a popular basic waveform, its primitive version falls short in sensing due to the inherent unregulated auto-correlation properties. Furthermore, the sensitivity to Doppler shift hinders its broader applications in dynamic scenarios. To address these issues, we propose a superposed index-modulated OFDM (S-IM-OFDM). The proposed scheme improves the sensing performance without excess power consumption by translating the energy efficiency of IM-OFDM onto sensing-oriented signals over OFDM. Also, it maintains excellent communication performance in time-varying channels by leveraging the sensed parameters to compensate for Doppler. Compared to conventional OFDM, the proposed S-IM-OFDM waveform exhibits better sensing capabilities and wider applicability in dynamic scenarios. Both theoretical analyses and simulations corroborate its dual benefits.

\end{abstract}


\IEEEpeerreviewmaketitle

\section{Introduction}

\IEEEPARstart{I}NTEGRATED sensing and communications (ISAC) has been receiving significant interests from both academia and industry\cite{cheng2022integrated}. Among the various studies conducted on ISAC, researchers have made significant efforts to develop the dual-functional waveform to perform both wireless communication and radio sensing simultaneously \cite{overview_ISAC_roadahead}. 
%Many design methods have been proposed to optimize the performance of dual-functional waveforms\cite{radar_cen_1,majorcom,joint2,joint3,hwang2008ofdm,OFDMbased_1,liuyongjun_super}. \cite{radar_cen_1,majorcom} embedded information into sensing-oriented radar signals at the cost of limited communication rate. \cite{joint2,joint3} jointly optimized the waveform but the complexity involved is relatively high. Therefore communication-centric approaches\cite{hwang2008ofdm,OFDMbased_1,liuyongjun_super}, where dual-functional waveforms built upon the OFDM signal consist of the most popular choice, are widely considered in practice.

%Dual-functional waveforms have the potential of reducing interference and improving efficiency in wireless networks \cite{overview1}. Many design methods have been proposed to optimize the performance of dual-functional waveforms, including communication-centric approaches\cite{commu_cen_1,commu_cen_2}, sensing-centric designs\cite{radar_cen_1,majorcom}, and joint ones  \cite{joint2,joint3}.

%In the communication-centric category, dual-functional waveforms built upon the OFDM signal consist of the most popular choice \cite{hwang2008ofdm,OFDMbased_1,liuyongjun1,liuyongjun_super}. 
%\vspace{-0.4cm}
Among the numerous designs, orthogonal frequency division multiplexing (OFDM) comes as a frequent candidate to enable dual functions without significantly altering the air interface \cite{hwang2008ofdm,OFDMbased_1,mimo_ofdm_isac,liuan_OFDMISAC}. Recall that OFDM exhibits two fundamental drawbacks in supporting sensing and communications. For the former, the sensing capability is unpredictable due to the unregulated sidelobes caused by the randomness of the transmitted information. For the latter, OFDM encounters performance degradation in time-varying channels due to inter-carrier interference (ICI). 
To enhance the sensing accuracy of OFDM, \cite{liuyongjun_super} employed 2D-multiple signal classification (MUSIC), a high-resolution estimation algorithm. To improve the theoretical performance of target detection, \cite{liuyongjun1} focused on optimizing the power allocation of OFDM subcarriers. Unfortunately, these methods cannot fundamentally tackle the issue of high sidelobes inherently from OFDM.
Towards the communication aspect, \cite{ICI_cancel_2} proposed self-cancellation schemes that cancel ICI across adjacent subcarriers, at the cost of the spectral efficiency. Another solution proposed by \cite{fan2021wideband} is a beamspace Doppler compensator, but it requires the assistance of massive antennas for beamspace projection.

%Towards the sensing aspect, what is the current effort/limitation?
%Towards the communicaiton aspect, what is the current effort/limitaiton?
%(This is where you can briefly summarize the existing work from two aspects). The current version only taps into the second aspect.

%In light of existing deficiencies, we propose an enhanced OFDM termed as superposed index-modulated OFDM (S-IM-OFDM), through the superposition of transparent sequence and IM-OFDM waveform. Specifically, IM-OFDM is able to convey information bits with reduced power consumption through deliberated design (see e.g., \cite{bacsar2013orthogonal,Shijian_precoded,wen_enhanced,van2018spread}), while the transparent sequence is overlaid to improve sensing capability by enhancing the auto-correlation property. Both components of S-IM-OFDM can be utilized for target sensing, thus fusing their individual estimate can boost the accuracy of range-velocity estimation. Moreover, the inherent sensing capability of S-IM-OFDM can be effectively employed to achieve proactive Doppler compensation, thereby substantially mitigating the impact of ICI in time-varying channels. Unlike the method proposed in \cite{csahin2021ofdm}, in which the radar signal is inserted into some of the inactive subcarriers, we utilize all subcarriers in IM-OFDM to render a smaller information loss and wider spectrum for sensing.

Recognizing these existing deficiencies, we propose an enhanced OFDM scheme termed as ``superposed index-modulated OFDM" (S-IM-OFDM). S-IM-OFDM combines a transparent sequence with the IM-OFDM waveform. IM-OFDM is known for conveying information bits with reduced power consumption through deliberate design (as seen in \cite{bacsar2013orthogonal,Shijian_precoded,wen_enhanced,van2018spread,guoshuaishuai}). It has also shown potential in applications for ISAC in \cite{csahin2021ofdm, wangxiangrong_IM_ISAC}. The transparent sequence is overlaid to enhance sensing capabilities by improving the auto-correlation property. Both components of S-IM-OFDM can be utilized for target sensing, and their combined estimates enhance the accuracy of range-velocity estimation. Additionally, the inherent sensing capability of S-IM-OFDM is effectively employed to proactively compensate for Doppler effects, significantly reducing the impact of ICI in time-varying channels. Unlike the method proposed in \cite{csahin2021ofdm}, where the radar signal is inserted into some of the inactive subcarriers, we utilize all subcarriers in IM-OFDM to ensure zero information loss and a broader spectrum for sensing. Compared to another IM-based solution in \cite{wangxiangrong_IM_ISAC}, where the information is embedded solely in the radar signals, our approach leverages both indices and symbols to convey bits, thus improving the achievable spectrum efficiency.

In addition to outlining the construction and operation of S-IM-OFDM, we conduct theoretical analysis to optimize its dual functionality. Simulations have confirmed the advantages of S-IM-OFDM over existing OFDM-related candidates. These advantages underscore the significant potential of S-IM-OFDM to address the shortcomings associated with traditional OFDM, making it a promising candidate for ISAC applications, especially in challenging time-varying channel conditions.

\newcommand{\RNum}[1]{\uppercase\expandafter{\romannumeral #1\relax}}

\section{Fundamentals of S-IM-OFDM}
In this section, we will describe the fundamentals of the S-IM-OFDM and illustrate how it serves for communication and sensing purposes.

\vspace{-0.2cm}
\subsection{Construction of S-IM-OFDM}
\subsubsection{Communication-Oriented Signal}
As can be inferred from the name, the core of S-IM-OFDM lies in IM-OFDM. Assume that $M$ subcarriers are available, with $\Delta f$ being the subcarrier interval and $T_{s}$ being the symbol duration. To ease implementation, these subcarriers are divided into $G$ groups, each consisting of $N_{g}=M/G$ subcarriers. In each group, $k$ out of $N_g$ subcarriers are activated to transmit modulated symbols equiprobably drawn from a specific constellation $\mathcal{S}$, as operated in \cite{bacsar2013orthogonal}. The indices of those activated subcarriers are stored in $\bm I_{g}$. During the $n$-th symbol $(n=1,\cdots, N_s)$, its communication-oriented signal is represented as $\bm x_{c}(n)=\left [x_{c,1}(n),\cdots,x_{c,M}(n)\right]^{T}$, where $x_{c,m}(n)\in \mathcal{S}$, if $m\in {\bm I_{1}\cup\bm I_{2}\cup\cdots \cup\bm I_{N_g}}$, and $x_{c,m}(n)=0$ otherwise. The total number of information bits per symbol is $G(\binom{N_{g}}{k}+k\log_{2}(\vert \mathcal{S}\vert))$, with $\binom{n}{k}$ denoting the number of $k$ combinations from a given set of $n$ elements, and $|\mathcal{S}|$ denotes the set's cardinality.
\subsubsection{Sensing-Oriented Signal}
The waveform superposed with IM-OFDM is a transparent pseudo sequence bearing excellent auto-correlation properties. This sequence, oriented to sensing, is represented as $\bm x_{s}(n)=\left [ x_{s,1}(n), \cdots,x_{s,M}(n)\right]^{T}$. Apparently, various options are available for $\bm x_{s}(n)$. Here we fix it as the well-known $m$-sequence for simplicity. It can be readily verified that $\mathbb{E}\left [ \bm x_{c}^{H}(n)\bm x_{s}(n) \right]=0$, indicating the two elements of S-IM-OFDM are statistically orthogonal.

\subsubsection{S-IM-OFDM}
To ensure the same transmission power as the case without superposition, the S-IM-OFDM is constructed as a linear combination of $\bm x_{c}$ and $\bm x_{s}$, adjusted by a power splitting ratio $\rho$ as follows:
\begin{equation}
\setlength{\abovedisplayskip}{2pt}
\setlength{\belowdisplayskip}{2pt}
    \bm x(n)=\sqrt{\rho} \bm x_{s}(n)+\sqrt{1-\rho}\bm x_{c}(n).
    \label{superposed signal baseband}
\end{equation}
 It is worth mentioning that the power splitting ratio $\rho$ is a system design parameter which could be unknown to the receiver.
\vspace{-0.24cm}

\subsection{Operations of S-IM-OFDM}

%\textcolor{red}{The S-IM-OFDM is transmitted and the information bits are decoded at the user to achieve the communication functionality. The echo of the S-IM-OFDM is reflected by the targets back to the transmitter for instant range and velocity estimation.}

%\textcolor{red}{The transmitter is equipped with a pair of antennas for signal transmitting and echo receiving respectively. The S-IM-OFDM achieves instant range and velocity estimation when the echo of the S-IM-OFDM is reflected by the targets back to the transmitter. With the estimation, the Doppler compensator is designed at the transmitter to actively mitigate the ICI. Finally, the information bits in S-IM-OFDM are decoded at the user. The operation diagram of the proposed S-IM-OFDM is shown in Fig.~\ref{fig:diagram}.}

%The transmitter is equipped with a pair of antennas, one for signal transmitting and the other for echo receiving. The S-IM-OFDM system achieves instant estimation of range and velocity by analyzing the echo of the S-IM-OFDM that is reflected back to the transmitter by the targets. The Doppler compensator is then designed at the transmitter to actively mitigate the effects of ICI. Finally, the user decodes the information bits in the S-IM-OFDM signal. 

Once putting S-IM-OFDM into dual-functional usage, the full operational steps include \emph{target sensing}, \emph{Doppler pre-compensation} and \emph{bits decoding}. To gain more intuition, the system diagram is shown in Fig. 1.

\begin{figure}[t]
  \vspace{-0.1cm}
  \setlength{\abovecaptionskip}{-0cm} %调整图片标题与图距离
  \setlength{\belowcaptionskip}{0.1cm} %调整图片标题与下文距离
  \centering
  \includegraphics[width=0.98\linewidth]{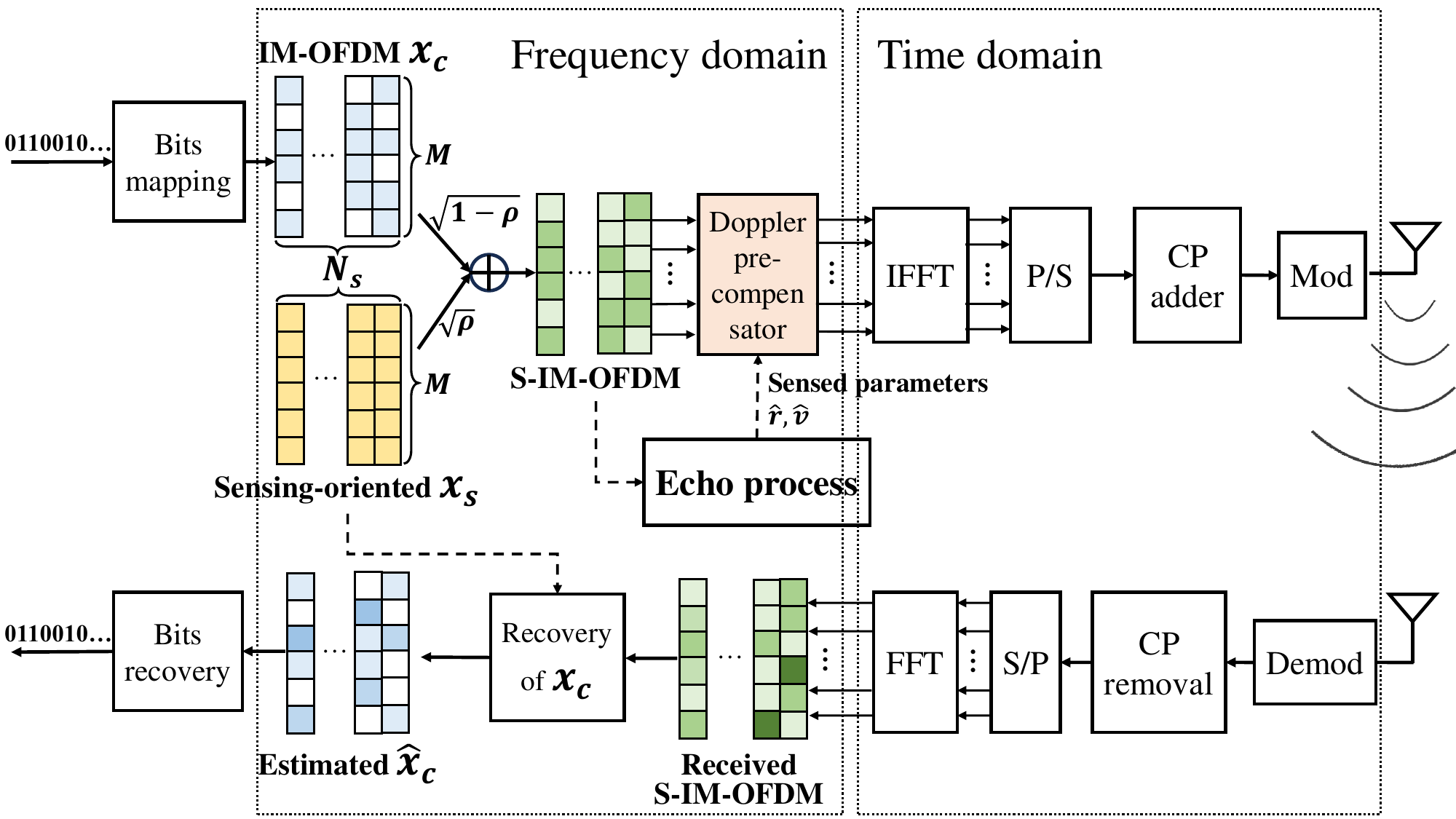}
  \caption{Systematic diagram of S-IM-OFDM transceivers.}
  \vspace{-0.4cm}
  \label{fig:diagram}
\end{figure}

\subsubsection{Target Sensing}\label{sec:sensing}
Since the transmitter has the complete knowledge of the  waveform aside from sensing-oriented $\bm x_{s}$, $\bm x_{c}$ itself can also assist in sensing \cite{csahin2021ofdm}. Thanks to the statistical orthogonality between $\bm x_{c}$ and $\bm x_{s}$, their processing can be implemented individually. The resulting two estimates can then be fused to reduce the error.

%For convenience, the received data rm,n is formed into a matrix, where each row represents an OFDM symbol and each column represents a subcarrier.

Assume the estimation is made over $N_s$ consecutive symbol durations. The continuous echo is first sampled, then goes to the CP removal, and finally is retrieved from the time-domain signal via discrete Fourier transform (DFT) operation. Let $\bm r$ and $\bm v$ be the targets' range and velocity, then the reflected signal in the frequency domain is expressed as 
%When using the transmitted signal for sensing, the first step is to obtain its reflected version in the frequency domain, \textcolor{blue}{which can be represented as
\begin{equation}
\setlength{\abovedisplayskip}{1pt}
\setlength{\belowdisplayskip}{1pt}
    \bm R_{s}=\bm G(\bm r,\bm v)\odot \bm X
    +\bm \xi_{s},
    \label{equ:echo}
\end{equation}
where $\odot$ denotes the Hadamard product. $\bm X$ and $\bm R_{s}\!\in \!\mathbb{C}^{M\times N_s}$ contain $N_s$ transmitted S-IM-OFDM symbols $\bm x(n)$ and their reflected copies. $G_{m,n}(\bm r,\bm v)=\sum_{p=1}^{P}\gamma_{p}e^{-jm\tau_{p}^{s}}e^{j2\pi nf_{p}^{s}}$ represents the target response. For the $p$-th target, $\gamma_{p}\sim \mathcal{CN}(0, \sigma_{\gamma}^{2})$ is the overall attenuation due to the propagation loss, scattering and radar-cross section, $\tau_{p}^{s}=4\pi \Delta fr_{p}/c_{v}$ and $f_{p}^{s}=2f_{c}v_{p}T_{s}/c_{v}$ are the normalized delay and Doppler shift of the $p$-th target respectively. $\bm \xi_{s} \sim\mathcal{CN}(0,\sigma_{s}^{2}\bm I)$ is the noise.
%\vspace{-0.5cm}
\begin{itemize}
\setlength{\abovedisplayskip}{2pt}
\setlength{\belowdisplayskip}{2pt}
    \item Processing over $\bm x_{s}$: By correlating $\bm R_s$ with $\bm X_{s}$ at the estimated $(\hat{\bm r},\hat{\bm v})$, the spectrum is obtained as $f(\hat{\bm r},\hat{\bm v})=\sum_{n=1}^{N_{s}}\sum_{m=1}^{M} R_{m,n}X_{s,m,n}e^{jm\hat{\tau}_{p}^{s}}e^{-j 2\pi n\hat{f}^{s}_{p}}$, 
%$f(\hat{\bm r},\hat{\bm v})=\text{sum}(\bm X_{s}\odot(\bm G^{H}(\hat{\bm r},\hat{\bm v})\odot \bm R_{s}))$.
%\begin{equation}
%\setlength{\abovedisplayskip}{2pt}
%\setlength{\belowdisplayskip}{2pt}
% f(\hat{\bm r},\hat{\bm v})=\sum_{n=1}^{N_{s}}\sum_{m=1}^{M} R_{s}_{m,n}X_{s}_{m,n}e^{jm\hat{\tau}_{p}}e^{-j n\hat{\omega}_{p}}
%\end{equation}
%\begin{equation}
%\setlength{\abovedisplayskip}{2pt}
%\setlength{\belowdisplayskip}{2pt}
%    f(\hat{\bm r},\hat{\bm v})=\text{sum}(\bm X_{s}\odot(\bm G^{H}(\hat{\bm r},\hat{\bm v})\odot \bm R_{s}))
%\end{equation}
%$\hat{\tau}_{p}= 4\pi \Delta f\hat{r}_{p}/c_{v}$ and $\hat{\omega}_{p}=2\pi f{c}\hat{v}_{p}T_{s}/c_{v}$
%$\text{sum}(\cdot)$ denotes the sum of all elements of a matrix
The estimates of the actual targets, $\left [ \hat{\bm r}_{1}, \hat{\bm v}_{1} \right ]$, are identified as the peak from the range-velocity power spectrum $f(\hat{\bm r},\hat{\bm v})$. %The variances of the estimation errors for range and velocity via $\bm x_s$ are denoted as $\sigma_{r,1}^{2}$ and $\sigma_{v,1}^{2}$.
 \item Processing over $\bm x_{c}$: Similar to \cite{xie2021performance}, the estimation process involves element-wise division, 2D-smoothing and obtaining 2D-spectrum using 2D-MUSIC algorithm. The estimated range and velocity are $\left [ \hat{\bm r}_{2}, \hat{\bm v}_{2} \right ]$. %with their respective error variances $\sigma_{r,2}^{2}$ and $\sigma_{v,2}^{2}$.
 \item Information fusion: With two sets of estimates, a linear fusion is applied as follows to reduce the estimation error:
     \begin{align}
         &\hat{\bm r}=w_{r}\hat{\bm r}_{1}+(1-w_r)\hat{\bm r}_{2},  \nonumber  \\ 
         &\hat{\bm v}=w_{v}\hat{\bm v}_{1}+(1-w_v)\hat{\bm v}_{2}.
     \end{align}
\vspace{-0.05cm}
The specific setting for linear coefficients $w_{r}$ and $w_{v}$ will be introduced later.
\end{itemize}

\subsubsection{Doppler Pre-Compensation}

In the presence of Doppler shifts, the discrete channel impulse response at the $d$-th delay tap ($d=0,1,\cdots, N_{d}-1$) is given by
\vspace{-0.1cm}
\begin{equation}
\setlength{\abovedisplayskip}{1pt}
\setlength{\belowdisplayskip}{1pt}
    h_{d}(m)=\sum_{p=1}^{P}\alpha_{p}\delta(d\frac{T_{s}}{M}-\tau_{p})e^{j2\pi f_{p} \frac{m}{M}T_{s}},
\vspace{-0.1cm}
\end{equation}
where $\alpha_{p}\sim \mathcal{CN}(0,\sigma_{\alpha}^{2})$, $\tau_p\sim\mathcal{U}[0,(N_{d}-1)T_{s})$ and $f_p=\frac{f_{c}v_{p}}{c_v}$ represent the complex gain, the delay, and the Doppler shift for the $p$-th path according to \cite{Doppler_channel_zhao}. Specifically, $\alpha_{p}$ can be estimated by methods in \cite{overview_ISAC_roadahead}, and $v_{p}$ is the normalized radial velocity. Through Fourier transform, the frequency-domain channel matrix $\bm H$ is expressed as
\vspace{-0.1cm}
\begin{equation}
\setlength{\abovedisplayskip}{1pt} 
\setlength{\belowdisplayskip}{1pt}
H_{i,j}=\frac{1}{M}\sum_{p=1}^{P}\alpha_{p}e^{-j\frac{2\pi\tau_{p}j}{T_{s}}}\frac{1-e^{j2\pi(f_{p}T_{s}-i+j)}}{1-e^{j2\pi(f_{p}T_{s}-i+j)/M}}. 
\vspace{-0.1cm}
\end{equation}
Using the inherent mono-static sensing capability of S-IM-OFDM, we can estimate $\tau_p$ and  $f_p$ according to the targets sensing method detailed in Section \ref{sec:sensing}, i.e., $\hat{\tau}_{p}=\hat{\tau}^{s}_{p}/2=\frac{\hat{r}_{p}}{c_{v}}$ and $\hat{f}_{p}=\hat{f}^{s}_{p}/2=\frac{f_{c}\hat{v}_{p}}{c_v}$. These estimates can then be leveraged to actively reconstruct the channel as $\widetilde{H}_{i,j}=\frac{1}{M}\sum_{p=1}^{P}\alpha_{p}e^{-j\frac{2\pi\hat{\tau}_{p}j}{T_{s}}}\frac{1-e^{j2\pi(\hat{f}_{p}T_{s}-i+j)}}{1-e^{j2\pi(\hat{f}_{p}T_{s}-i+j)/M}}$. Following \cite{equal_tcom} and \cite{ICI_equ}, we place a Doppler pre-compensator, $\bm U = \frac{\widetilde{\bm H}^{-1}}{\| \widetilde{\bm H}^{-1}\|_{F}}$, before IFFT to eliminate ICI through frequency-domain filtering, while satisfying the total power constraint, as illustrated in Fig.~\ref{fig:diagram}.
%compensate for the time-varying channels. Following \cite{ICI_equ}, we place a Doppler pre-compensator, $\bm U = \frac{\widetilde{\bm H}^{-1}}{\| \widetilde{\bm H}^{-1}\|_{F}}$, before IFFT to eliminate ICI through frequency-domain filtering, while satisfying the total power constraint, as illustrated in Fig.~\ref{fig:diagram}. Note that the mono-static sensing capability of S-IM-OFDM makes it possible to actively reconstruct the channel as $\widetilde{H}_{i,j}=\sum_{m=1}^{M}\sum_{p=1}^{P}\frac{1}{M}\alpha_{p}e^{j\frac{2\pi}{M}\left[ i(m-\hat \tau_p)+m\frac{\hat{\omega}_{p}}{2\pi\Delta f}-mj\right]}$, with the estimated parameters $\hat \tau_p$ and $\hat \omega_p$. It is worth mentioning that the delay and Doppler frequency shift parameters in the estimated channel are half of mono-static estimates from S-IM-OFDM, i.e., $\tau_{p}=$ and $\omega_{p}=$.
Therefore, the received signal in frequency domain after channel propagation can be
\setlength{\abovedisplayskip}{1pt}
\setlength{\belowdisplayskip}{1pt}
%\begin{align}
%    \bm y&=\sqrt{P_{t}}\bm H \bm U\bm x+\bm \xi=\frac{\sqrt{P_{t}}}{\| \widetilde{\bm H}^{-1}\|_{F}}\bm H \widetilde{\bm H}^{-1} \bm x+\bm \xi \nonumber\\
%    &=\!\frac{\sqrt{P_{t}}}{\| \widetilde{\bm H}^{-1}\!\|_{F}}\!(\bm I\!+\!\bm \Delta_{H}\!\bm H^{-1})^{-1}\!\bm x\!+\!\bm \xi\!=\!\frac{\sqrt{P_{t}}}{\| \widetilde{\bm H}^{-1}\!\|_{F}}\!\overline{\bm H}\bm x\!+\!\bm \xi,
%\end{align}
\begin{align}
    \bm y&=\sqrt{P_{t}}\bm H \bm U\bm x+\bm \xi=\!\frac{\sqrt{P_{t}}}{\| \widetilde{\bm H}^{-1}\!\|_{F}}\!\overline{\bm H}\bm x\!+\!\bm \xi_{c},
\end{align}
where $P_t$ is transmit power, $\xi_{c}\sim \mathcal{CN}(0, \sigma_{c}^{2})$ is the received noise, $\overline{\bm H}=(\bm I\!+\!\bm \Delta_{H}\bm H^{-1})^{-1}$ is the equivalent channel in frequency domain for ease of notation, and $\bm \Delta_{H}=\widetilde{\bm H}-\bm H$ denotes the estimation error matrix. Without causing ambiguity, we drop the symbol index $n$ in detection design.

\subsubsection{Bits Decoding}
At the receiver, the transmitted bits can be decoded by subtracting the interference from $\bm x_{s}$. The detailed steps for recovering the communication signal are as follows:
\begin{itemize}
    \item Estimating the power ratio: 
    \begin{equation}
        \hat{\rho}=\frac{\|\bm H^{-1}\|_{F}^{2}\|\mathbb{E}\{\bm y\bm x_{s}^{H}\}\|_{F}^{2}}{MP_{t}}.
    \label{equ:rho}
    \end{equation}
    %$\hat{\rho}=\frac{\|\bm H^{-1}\|_{F}^{2}\|\mathbb{E}\{\bm y\bm x_{s}^{H}\}\|_{F}^{2}}{MP_{t}}$.
    \item Subtracting the reconstructed sensing signal: 
    \begin{equation}
        \tilde{\bm x}_{c}=\frac{\frac{\|\bm H^{-1}\|_{F}}{\sqrt{P_{t}}}\bm y-\sqrt{\hat{\rho}}\bm x_{s}
    }{\sqrt{1-\hat{\rho}}}.
    \label{substraction}
    \end{equation}
    \item Applying the maximum-likelihood (ML) detection within each group:
\begin{equation}
\setlength{\abovedisplayskip}{2pt}
\setlength{\belowdisplayskip}{2pt}
    (\hat{\bm I}_{g},\hat{\bm x}_{g} )=\mathop{\arg\min}\limits_{\bm I_{g}, \bm x_{g}}\|\bm x_{c,\text{IM}}(\bm I_{g}, \bm x_{g})-\tilde{\bm x}_{c,g}\|_{2}^{2}.\label{con:jietiao}
\vspace{-0.1cm}
\end{equation}
where $\hat{\bm I}_{g}$ and $\hat{\bm x}_{g}$ are the estimated set of activated subcarriers and information symbols in the $g$-th group. $\bm x_{c,\text{IM}}(\bm I_{g}, \bm x_{g})$ is the index-modulated communication symbol in the $g$-th group with $\bm x_{g}$ at the subcarriers $\bm I_{g}$.

%$\hat{\bm I}_{g}$ denote the possible set and the estimated set of activated subcarriers in the $g$-th group, $\bm x_{g}$ and $\hat{\bm x}_{g}$ denote the possible symbols and the estimated symbols at the corresponding subcarriers in the $g$-th group, and $\bm x_{c,\text{IM}}(\bm I_{g}, \bm x_{g})$ is the index-modulated communication symbol with $\bm x_{g}$ at the active subcarriers $\bm I_{g}$. With the detected indices and symbols, the transmitted bits are recovered at the user.
\end{itemize}

\section{Analysis of S-IM-OFDM}
In this section, we quantify the performance of communication and sensing with S-IM-OFDM, aiming to offer insightful guidance in practical design.

\vspace{-0.3cm}
\subsection{Communication Performance}

As a key component of S-IM-OFDM, IM-OFDM is able to boost a higher coding gain over OFDM, making it possible to shift some power to the sensing part under the same bit error rate (BER) requirement. However, this allocation is not arbitrary because a larger residue arising from mediocre extraction of $\bm x_{s}$ may render a worse BER performance.

%interfere the bits decoding. For this reason, it is crucial to determine the proper power allocation for optimized dual-functional performance.

%\textcolor{red}{We define $\Gamma(\mathcal{A},\gamma)$ as the required signal-to-noise ratio (SNR) for a certain wavefrom $\mathcal{A}$ to achieve the bit error rate (BER) as $\gamma$.} According to \cite{wen_enhanced}, the coding gain of IM-OFDM in the proposed waveform over OFDM at $\text{BER}=\gamma$ can be represented as 
%\begin{equation}
%\setlength{\abovedisplayskip}{2.5pt}
%\setlength{\belowdisplayskip}{2.5pt}
%   G_{c}=\frac{\Gamma(\text{OFDM},\gamma)-\Gamma(\text{IM-OFDM},\gamma)}{\Gamma(\text{OFDM},\gamma)}
   %\label{equ:coding gain}
%\end{equation}

Substituting the power estimation $\hat{\rho}$ defined in Eq.~(\ref{equ:rho}) into Eq.~(\ref{substraction}) yields
\setlength{\abovedisplayskip}{2pt}
\setlength{\belowdisplayskip}{2pt}
\begin{align}
    \tilde{\bm x}_{c}\!=\!\overline{\bm H}\bm x_{c}
    \!+\!\frac{\sqrt{\rho}}{\sqrt{1\!-\!\rho}}(\overline{\bm H}\!-\!\frac{\|\overline{\bm H}\|_{F}}{\sqrt{M}}\bm I)\bm x_{s}\!+\!\frac{\| \bm H^{-1}\|_{F}}{\sqrt{1\!-\!\rho}\sqrt{P_{t}}} \bm\xi.
\end{align}
The signal-to-interference-plus-noise ratio (SINR) at the $m$-th subcarrier can be computed as 
\begin{equation}
\setlength{\abovedisplayskip}{2pt}
\setlength{\belowdisplayskip}{2.5pt}
    \text{SINR}_{m}=\frac{1}{\Omega_{m}(\overline{\bm H})\!+\! \frac{\| \bm H^{-1}\|^{2}_{F}\sigma_{\xi}^{2}}{(1-\rho)\vert\overline{H}_{m,m}\vert^{2}P_t}},
   \label{equ:SINR}
\end{equation}
where $\Omega_{m}(\overline{\bm H})=\frac{\sum_{j\neq m}\vert\overline{H}_{m,j}\vert^{2}+\frac{\rho}{1-\rho}(\overline{H}_{m,m}-\frac{\| \overline{\bm H}\|_{F}}{\sqrt{M}})^{2}}{\vert\overline{H}_{m,m}\vert^{2}}$ depends on both the instantaneous channel and the precision of compensation.
Properly setting $\rho$ can effectively suppress the noise brought by $\bm x_{s}$ while enhancing the effectiveness of Doppler compensator, thus maximizing SINR.
Therefore, to optimize the communication performance in terms of BER, the power splitting ratio $\rho$ should maximize the minimal SINR across all subcarriers, i.e.,
%To minimize the BER at the receivers, the optimal power splitting ratio $\rho$ should maximize the minimal SINR across all subcarriers, which is given by
%The optimal power allocation ratio of sensing-oriented signal to S-IM-OFDM symbols, $\rho$, that yields the best communication reliability, namely the lowest bit error rate(BER), is
\begin{equation}
\setlength{\abovedisplayskip}{2pt}
\setlength{\belowdisplayskip}{2pt}
    \rho^{*}=\mathop{\arg\max}\limits_{\rho}\{\min_{m} \text{SINR}_{m}\}.
    \label{equ:rho_com_opt}
\vspace{-0.1cm}
\end{equation}
\vspace{-0.1cm}

However, the communication-optimal power splitting ratio in Eq.~(\ref{equ:rho_com_opt}) lacks consideration of sensing tasks, and may not be optimal in dual-functional applications. Reasonably setting $\rho$ can provide greater freedom for the sensing part while ensuring a performance gain in communication over OFDM.
%and is not necessarily optimal in practical dual-functional applications. Reasonably setting $\rho$ can give higher freedom to the sensing part, while guaranteeing the communication performance gain over traditional OFDM.
Let's define $G_{\text{c}}(\gamma)$ (dB) as the signal-to-noise ratio (SNR) gap when IM-OFDM achieves a similar BER $\gamma$ as that of OFDM.
\vspace{-0.5cm}
%This definition allows us to draw an important conclusion concerning the power splitting ratio after some derivations:}
\begin{proposition}
    To ensure S-IM-OFDM is no worse than OFDM in terms of communication performance, the power splitting ratio $\rho$ is upper bounded by
    \begin{equation}
\setlength{\abovedisplayskip}{2pt}
\setlength{\belowdisplayskip}{1pt}
    \rho_{\text{max}}=1-\max_{m}\left\{\!\frac{\| \bm H^{-1}\|^{2}_{F}\sigma_{\xi}^{2}}{10^{\frac{G_{\text{c}}}{10}}(\Omega_{m}(\overline{\bm H})+\frac{\| \bm H^{-1}\|^{2}_{F}\sigma_{\xi}^{2}}{P_{t}^{2}\vert \overline{H}_{m,m}\vert^{2}})\vert \overline{H}_{m,m}\vert^{2} P_{t}^{2}}\!\right\}.
    %\nonumber
    \label{equ:UB_time_varying}
\end{equation}
\end{proposition}
\vspace{-0.3cm}
\begin{remark}
\vspace{-0.5cm}
%Upon satisfying Proposition 1, the minimum SINR of S-IM-OFDM may exceed the maximum SINR of the OFDM benchmark, thereby enhancing communication reliability.
In static channels, $\Omega_{m}(\overline{\bm H})=0$, giving rise to $\rho_{max}=1-10^{-\frac{G_{c}}{10}}$. As a result, SINR in Eq.~(11) becomes monotocally increasing with $\rho$, implying a larger $\rho$ is favored by communication at the cost of declined sensing capability.
\end{remark}

%\vspace{-0.2cm}
\vspace{-0.5cm}
\subsection{Sensing Performance}

Assume the two independent velocity estimates via $\bm x_{s}$ and $\bm x_{c}$ are linearly fused as Eq.~(3). 
If there is no prior information about the target, we simply do the averaging ($\omega=0.5$). In many cases, the prior is available based on the historical estimates from the dual-functional waveform itself \cite{liufan_V2I_bayesian} or the message uploaded by other sensing nodes in sensor networks \cite{localization_nodes_tvt}.
%However, the dual-functional BS can obtain prior information about the target's position from its historical estimates \cite{liufan_V2I_bayesian} or message uploaded by other nodes \cite{localization_nodes_tvt}, 
Even though the prior might be noisy or inaccurate, we could still utilize the linear fusion based on the criterion of minimum mean squared error (MMSE) in \cite{lo1999maximum}, with the weighting coefficient being
\vspace{-0.05cm}
\begin{equation}
    w_{v}=\frac{\sigma_{v,1}^{2}\sigma_{v,2}^{2}}{\sigma_{v,1}^{2}(\sigma_{v,1}^{2}+\sigma_{v,2}^{2})},
\end{equation}
%$w_{v}^{\ast}=\frac{\sigma_{v,1}^{2}\sigma_{v,2}^{2}}{\sigma_{v,1}^{2}(\sigma_{v,1}^{2}+\sigma_{v,2}^{2})}$, 
where $\sigma_{v,1}$, $\sigma_{v,2}$ denote the variances of estimation errors for velocity via $\bm x_{s}$ and $\bm x_{c}$ relative to the reference point, respectively. 
The range estimation can be handled similarly thus being omitted here.

%The ideal sensing performance bound can be achieved based on the criterion of minimum mean squared error (MMSE) in \cite{lo1999maximum}, i.e., $w_{v}^{\ast}=\frac{\sigma_{v,1}^{2}\sigma_{v,2}^{2}}{\sigma_{v,1}^{2}(\sigma_{v,1}^{2}+\sigma_{v,2}^{2})}$, where $\sigma_{v,1}$, $\sigma_{v,2}$ denote the variances of estimation errors for velocity via $\bm x_{s}$ and $\bm x_{c}$ respectively. %Accordingly the ultimate variance of velocity error becomes $\frac{\sigma_{v,1}^{2}\sigma_{v,2}^{2}}{\sigma_{v,1}^{2}+\sigma_{v,2}^{2}}$.

%When the ground truth can not be obtained online, $\sigma_{v,1}$, $\sigma_{v,2}$ become inaccessible. In such cases, we assign weights to the two separate estimates in proportion to their individual confidences, which are indicated by their auto-correlation traits：

%The range estimation can be handled similarly thus being omitted here.

%% result 1
\begin{figure*}[t]
  \vspace{-0.5cm}
  \setlength{\abovecaptionskip}{0cm} %调整图片标题与图距离
  \setlength{\belowcaptionskip}{-1cm} %调整图片标题与下文距离
  \centering
  \subfigure[range estimation\label{fig:RMSE_r}]
    {\includegraphics[width=0.35\linewidth]{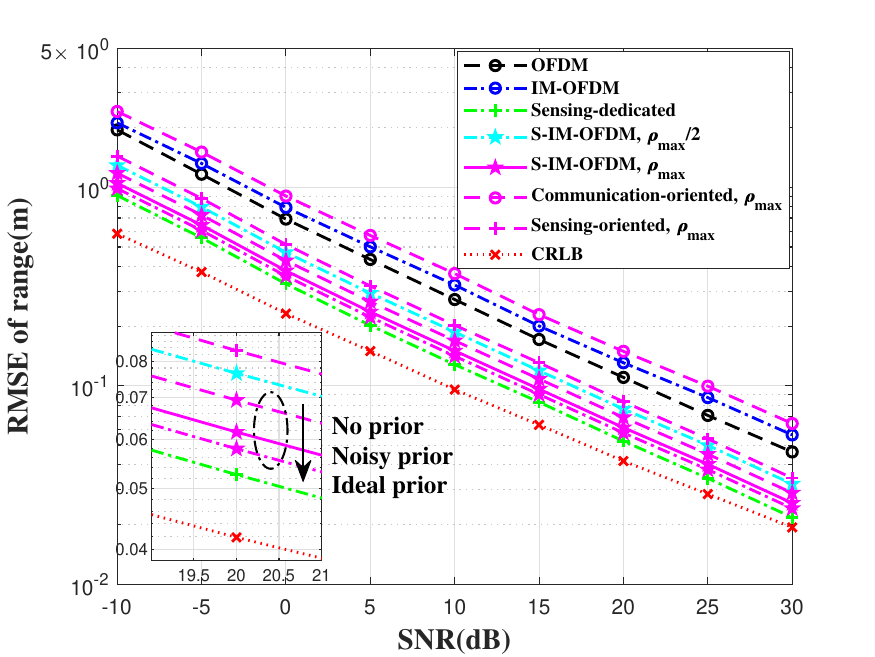}}
    \hspace{2.7cm}
  \subfigure[velocity estimation\label{fig:RMSE_v}]
    {\includegraphics[width=0.35\linewidth]{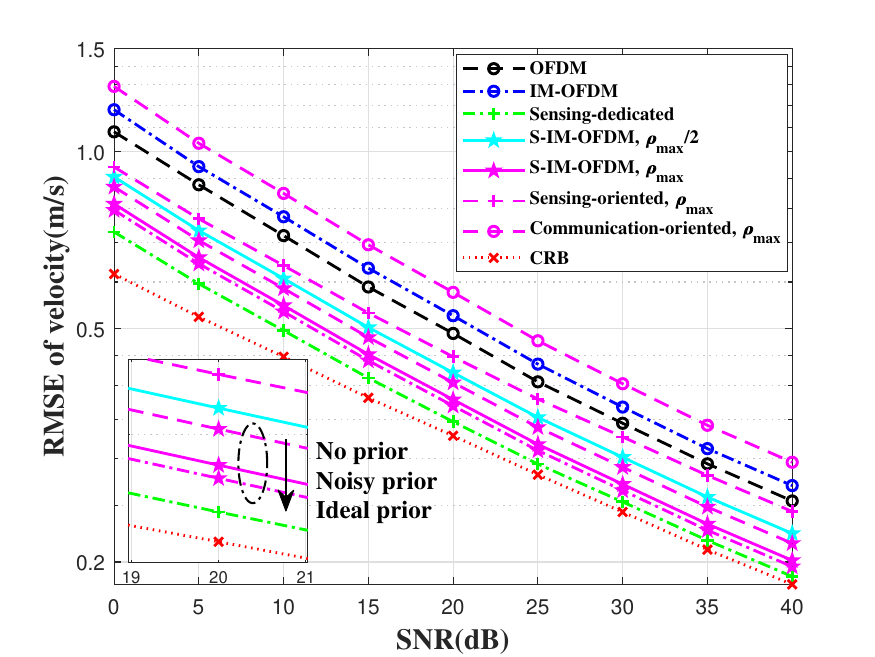}}
  \caption{RMSE comparisons among S-IM-OFDM, OFDM, IM-OFDM and sensing-dedicated waveforms.}
  \vspace{-0.2cm}
  \label{fig:sensing side}
\end{figure*}

We then derive the Cramér–Rao Lower Bound (CRLB) as a benchmark to evaluate this linear fusion estimator. Based on the Eq.~(2), the log-likelihood function of the received signal can be represented as 
\begin{align}
    \ln{p(\bm R_{s}\vert \bm r, \bm v)}\!=\!&-\!\frac{1}{\sigma_{s}^{2}}\!\sum_{n} \!\sum_{m}\vert R_{m,n}\!-\!G_{m,n}(\bm r,\bm v)X_{m,n}\vert^{2}\vspace{-1pt}\\ \nonumber
    \vspace{-0.4cm}
    &-\ln{(\pi^{MN_{s}}\sigma_{s}^{2MN_{s}})},
\end{align}
and the corresponding Fisher information matrix is $\bm J=\text{diag}(\bm J_{1}, \cdots, \bm J_{p}, \cdots, \bm J_{P})$,
%\begin{equation}
%    \bm J=\text{diag}(\bm J_{1}, \cdots, \bm J_{p}, \cdots, \bm J_{P}),
%\end{equation}
with
%\begin{equation}
%\setlength{\abovedisplayskip}{2pt}
%\setlength{\belowdisplayskip}{2pt}
%    \bm J_{p}=\begin{bmatrix} -\mathbb{E}\{\frac{\partial^{2}L}{\partial r_{p}^{2}}\} & -\mathbb{E}\{\frac{\partial^{2}L}{\partial r_{p} \partial v_{p}}\} \\ -\mathbb{E}\{\frac{\partial^{2}L}{\partial v_{p} \partial r_{p}}\} & -\mathbb{E}\{\frac{\partial^{2}L}{\partial v_{p}^{2}}\}\end{bmatrix}
%    \nonumber
%\end{equation}
\begin{align}
      [\bm J_{p}]_{1,1}&=-\mathbb{E}\left \{\frac{\partial^{2}\ln{p(\bm R_{s}\vert \bm r, \bm v)}}{\partial r_{p}^{2}}\right\} \nonumber \\
    &=\!\frac{j2\gamma_{p}P_{t}}{\sigma_{s}^{2}c_{v}^{2}}\!\sum_{n}\sum_{m}\!m^{2}\vert X_{m,n}\vert^{2}e^{j2\pi nf_{p}-\frac{j2mr_{p}}{c_v}} ,\nonumber \\
     [\bm J_{p}]_{1,2}&=[\bm J_{p}]_{2,1}=-\mathbb{E}\left\{\frac{\partial^{2}\ln{p(\bm R_{s}\vert \bm r, \bm v)}}{\partial r_{p} \partial v_{p}}\right\} \nonumber \\
     &=\!-\frac{4\pi\gamma_{p}f_{c}T_{s}P_{t}}{\sigma_{s}^{2}c_{v}^{2}}\!\sum_{n}\!\sum_{m}\!mn\vert X_{m,n}\vert^{2}e^{\frac{j2n\pi \!f_{c}T_{s}\!v_{p}}{c_{v}}-\frac{\!j2mr_{p}}{c_v}}, \nonumber \\
     [\bm J_{p}]_{2,2}&=-\mathbb{E}\left\{\frac{\partial^{2}\ln{p(\bm R_{s}\vert \bm r, \bm v)}}{\partial v_{p}^{2}}\right\} \nonumber \\
     &=\!\frac{4\pi^{2}\gamma_{p}f_{c}^{2}T_{s}^{2}P_{t}}{\sigma_{s}^{2}c_{v}^{2}}\sum_{n}\sum_{m}\!n^{2}\vert X_{m,n}\vert^{2}e^{\!\frac{j2n\pi \!f_{c}\!T_{s}\!v_{p}}{c_{v}}\!-\!\frac{j2m\!r_{p}}{c_v}}.\nonumber
\end{align}
Taking the inverse of $\bm J_{p}$, the CRLB of the range and velocity estimates are $\text{CRLB}(r_{p})=[\bm J_{p}^{-1}]_{1,1}$, $\text{CRLB}(v_{p})=[\bm J_{p}^{-1}]_{2,2}$.
%\begin{equation}
%\setlength{\abovedisplayskip}{2pt}
%\setlength{\belowdisplayskip}{2pt}
%     \begin{align}
%         &\text{CRLB}(r_{p})=[\bm J_{p}^{-1}]_{1,1},   \\ 
%         &\text{CRLB}(v_{p})=[\bm J_{p}^{-1}]_{2,2}.
%     \end{align}
%\end{equation}

\section{Simulation Results and Analysis}
In this section, we verify the effectiveness of the proposed scheme through numerical simulations. 
%Firstly, we analyze the root mean square error (RMSE) of estimations for range and velocity and compare them with those of OFDM and IM-OFDM. Subsequently, we present a comparison of BER performance of the proposed scheme with that of OFDM and IM-OFDM in both static and time-varying channels. 
The simulations are conducted using the following settings: carrier frequency $f_{c}=2.5$ GHz, $M=256$ subcarriers, $N_{g}=8$, $N_s=32$, a duration of the cyclic prefix $T_{g}=5\times 10^{-6}$~s and one symbol $T_{s}=6.67\times 10^{-5}$~s,
%a cyclic prefix duration of $T_{g}=5\times 10^{-6}$~s, a symbol duration of $T_{s}=6.67\times 10^{-5}$~s, 
and a subcarrier spacing of $\Delta f=15$ kHz. OFDM with binary phase shift keying (BPSK) is set as the baseline. In IM-OFDM, we use QPSK with 2 active subcarriers per group to maintain the same spectrum efficiency. The SNR is defined as the ratio of the power per bit to the noise power, i.e., $\frac{P_t}{b\sigma^{2}_{\xi}T_{s}}$, considering the normalized power of S-IM-OFDM symbols and the compensator, and $\sigma^{2}_{s}=\sigma^{2}_{c}=0$ dBm.
%Unless specified otherwise, the experiments consider Rician fading with a factor of $K = 2$.

We initially assess the sensing accuracy of S-IM-OFDM across various power splitting ratios and draw a comparison with OFDM and IM-OFDM, as illustrated in Figure 2. 
Without loss of generality, the range and the radial velocity parameters of $P=4$ targets are set as (15 m, 15 m/s), (30 m, 5 m/s), (45 m, 10 m/s), (80 m, 10 m/s) respectively, with $\sigma_{\gamma}=1$.
Employing linear fusion with proper weighting coefficients effectively reduces the root mean square error (RMSE) of S-IM-OFDM, surpassing the accuracy achieved by individual components and approaching the ideal performance bound from ideal prior. S-IM-OFDM consistently outperforms both OFDM and IM-OFDM in terms of sensing accuracy, all while operating at the same power consumption. This superiority can be attributed to the remarkable auto-correlation properties of sensing-oriented m-sequences, as well as the effectiveness of fusion estimation. Furthermore, as the power splitting ratio $\rho$ increases, the sensing capability is harnessed to a greater extent, resulting in a substantial reduction in error.
%It is worth emphasizing that the precise estimation of range-velocity, crucial for accurate Doppler pre-compensation, relies on S-IM-OFDM.

\begin{figure}[t]
  \vspace{-0.63cm}
  \setlength{\abovecaptionskip}{-0cm} %调整图片标题与图距离
  \setlength{\belowcaptionskip}{-1cm} %调整图片标题与下文距离
  \centering
  \includegraphics[width=0.7\linewidth]{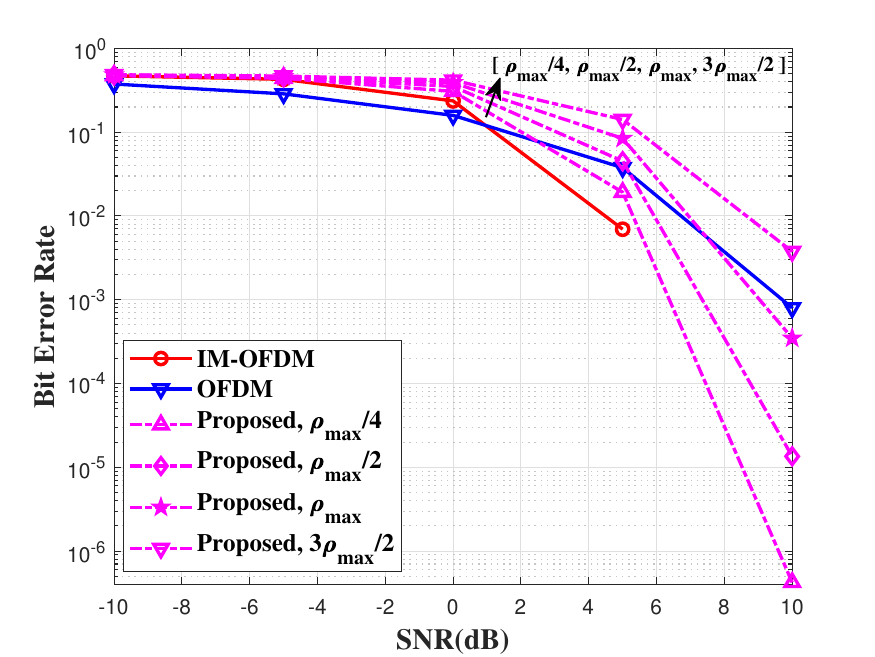}
  \caption{BER comparisons among S-IM-OFDM, IM-OFDM and OFDM waveforms in static channels.}
  \label{fig:communication_static}
  \vspace{-0.5cm}
\end{figure}

%In Fig.~3, we examine the communication performance in static channels. \textcolor{magenta}{The experiments consider Rician fading channels with $\sigma_{\alpha}=1$, $N_{d}=128$, $f_{p}=0$ and the Rice factor of $K = 2$.} In this setup, IM-OFDM achieves approximately a 5 dB gain compared to OFDM, thus being treated as the ideal benchmark. When SNR is below 5 dB, S-IM-OFDM is inferior to OFDM due to the imperfect orthogonality between the sensing-oriented signal and IM-OFDM. Increasing  SNR can enhance the orthogonality, thus preserving the coding advantage of S-IM-OFDM over OFDM. We also demonstrate the impact of the power allocation ratio $\rho$, revealing an inherent trade-off between sensing capability and communication reliability. A higher sensing-oriented power ratio consistently performs worse in terms of BER compared to a lower ratio, as it offers no coding benefits but introduces more noise. When $\rho$ exceeds the upper bound, the performance becomes even worse than that of OFDM.

In Fig.~3, we examine the communication performance in static channels. The experiments adopt Rician fading channels with $N_{d}=16$, $f_{p}=0$, the Rice factor $K = 2$ for the line-of-sight (LoS) path, and $\sigma_{\alpha}=1$ in non-line-of-sight (NLoS) paths. In this setup, IM-OFDM achieves approximately a 5 dB gain compared to OFDM, thus being treated as the ideal benchmark. When SNR is below 5 dB, S-IM-OFDM is inferior to OFDM due to the imperfect orthogonality between the sensing-oriented signal and IM-OFDM. Increasing SNR can enhance the orthogonality, thus preserving the coding advantage of S-IM-OFDM over OFDM. We also reveal an inherent trade-off between sensing and communication by demonstrating the impact of $\rho$. A higher $\rho$ consistently performs worse in terms of BER compared to a lower $\rho$, as it offers no coding benefits but introduces more noise. When $\rho$ exceeds the upper bound, the performance becomes even worse than that of OFDM.

%Subsequently, we validate the effectiveness of S-IM-OFDM in time-varying channels in Fig. 4. \textcolor{magenta}{The Doppler effect is incorporated into simulations with $\sigma_{D}^{2}=100$.} In conventional OFDM and IM-OFDM schemes, BER exhibits a floor in the high SNR region, due to severe ICI. Without compensation, the performance of S-IM-OFDM may be inferior to that of OFDM, mainly because of power allocation to the sensing-oriented part and the errors in index detection. Fortunately, through Doppler pre-compensation inherently brought by itself, S-IM-OFDM can achieve a lower BER than that of OFDM under high SNR, bridging the detection loss caused by the time-varying channel. The information fusion in S-IM-OFDM provides higher accuracy, ensuring more precise Doppler pre-compensation than that by OFDM. Additionally, an appropriate power splitting ratio $\rho$ can alleviate performance deterioration. As a comparison, insufficient precision in compensation occurs when $\rho$ is too low, while excessive background noise for communication arises when $\rho$ is too high. These phenomena align with the theoretical analysis.

Subsequently, we validate the effectiveness of S-IM-OFDM in time-varying channels in Fig. 4. The Doppler effect is incorporated into simulations with the normalized radial velocity (in m/s) $v_{p}\sim\mathcal{N}(0,100)$. In conventional OFDM and IM-OFDM schemes, BER exhibits a floor in the high SNR region, due to severe ICI. 
%Without compensation, the performance of S-IM-OFDM may be inferior to that of OFDM, mainly because of power allocation to the sensing-oriented part and the errors in index detection. 
Fortunately, through Doppler pre-compensation inherently brought by itself, S-IM-OFDM can achieve a lower BER than that of OFDM under high SNR, bridging the detection loss caused by the time-varying channel. The information fusion in S-IM-OFDM provides higher accuracy, ensuring more precise Doppler pre-compensation than that by OFDM. Additionally, an appropriate $\rho$ can alleviate performance deterioration. As a comparison, insufficient precision in compensation occurs when $\rho$ is too low, while excessive background noise for communication arises when $\rho$ is too high. These phenomena align with the theoretical analysis.

%% time-varying channel
\begin{figure}[t]
  \vspace{-0.5cm}
  \setlength{\abovecaptionskip}{-0cm} %调整图片标题与图距离
  \setlength{\belowcaptionskip}{-10cm} %调整图片标题与下文距离
  \centering
  \includegraphics[width=0.7\linewidth]{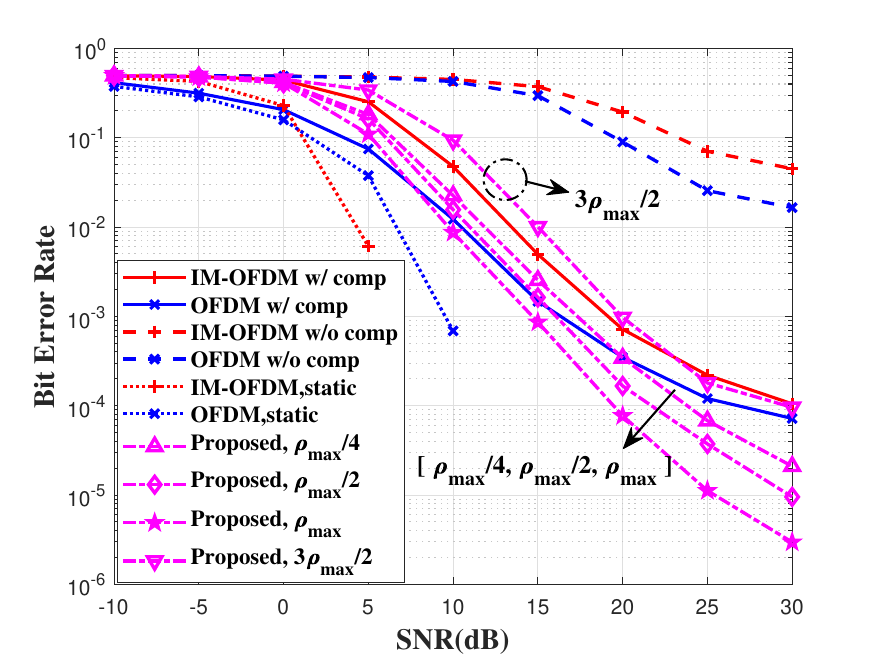}
  \caption{BER comparisons among S-IM-OFDM, IM-OFDM and OFDM waveform in time-varying channels.}
  \vspace{-0.5cm}
  \label{fig:doppler}
\end{figure}

In Fig.~5, we further test the communication performance by varying $P$. The system excels in maintaining a low BER in LoS channels ($P=1$). This success can be attributed to its precise estimation of individual propagation paths and efficient compensation for Doppler effects.
However, when dealing with an increasing number of paths in Rician channels, the system faces challenges in compensating for multiple delays and Doppler shifts, resulting in a degradation in performance, specifically in terms of BER. In comparison, in environments with rich scattering, such as Rayleigh channels, the system's compensation capabilities are further reduced. This leads to an even higher BER, although it still outperforms conventional OFDM. As a result, S-IM-OFDM demonstrates significant advantages, particularly in channels with prominent dominant paths. This implies the technique could be of great potential to doubly-selective mmWave channels \cite{doubly_selective} where few scatterers exist. The extension to mmWave massive MIMO will be left as the future work.

\section{Concluding remarks}

%shijian's version:
This paper introduces S-IM-OFDM, a novel dual-functional waveform tailored for ISAC applications. It effectively addresses OFDM's limitations by harnessing the energy efficiency of communication-oriented index modulation and the superior auto-correlation properties of sensing-oriented transparent sequences. The sensing and communication performance is flexibly adjusted via the power splitting ratio. Extensive simulations substantiate S-IM-OFDM's dual advantages. From a sensing perspective, it outperforms OFDM in range-velocity estimation accuracy. In the realm of communications, S-IM-OFDM consistently maintains a substantial coding gain over OFDM, especially in time-varying channels.

%By leveraging the energy efficiency of IM-OFDM and the superior auto-correlation properties of sensing-oriented transparent sequences, S-IM-OFDM can significantly improve the sensing accuracy. Additionally, the sensed information can inherently help overcome the deterioration of the communication performance in time-varying channels. 

%\appendices
%\section{Proof of the Proposition 1}
%Appendix 1 text goes here.

%\section{Proof of the Proposition 2}
%Appendix 2 text goes here.

% you can choose not to have a title for an appendix
% if you want by leaving the argument blank

% use section* for acknowledgment
%\section*{Acknowledgment}

%The authors would like to thank...

% Can use something like this to put references on a page
% by themselves when using endfloat and the captionsoff option.
\ifCLASSOPTIONcaptionsoff
  \newpage
\fi
\bibliographystyle{IEEEtran}
\bibliography{IEEEabrv,myrefs}
%% different channel conditions
\begin{figure}[t]
  \vspace{-0.5cm}
  \setlength{\abovecaptionskip}{-0cm} %调整图片标题与图距离
  \setlength{\belowcaptionskip}{-10cm} %调整图片标题与下文距离
  \centering
  \includegraphics[width=0.7\linewidth]{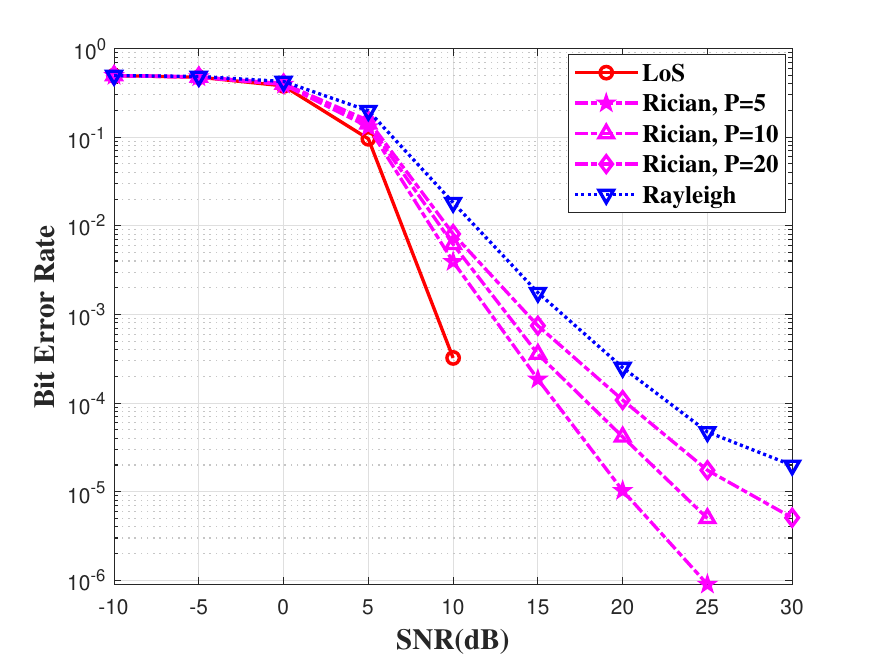}
  \caption{BER comparisons among different channel conditions.}
  \vspace{-0.5cm}
  \label{fig:channels}
\end{figure}

\end{document}